\documentclass[runningheads]{llncs}
\usepackage{graphicx}
\usepackage{rotating}
\usepackage{float}  
\usepackage[
monochrome
]{xcolor}
\usepackage[center]{caption}
\usepackage{multirow}
\usepackage{amsmath}
\usepackage{colortbl}
\usepackage{xcolor}
\begin{document}
\title{Wearable-based Fair and Accurate Pain Assessment Using Multi-Attribute Fairness Loss in Convolutional Neural Networks}
%
%
%
\author{
Yidong Zhu\inst{1} \and
Shao-Hsien Liu\inst{2} \and
Mohammad Arif Ul Alam\inst{1,2,3}}

\authorrunning{Zhu et al.}
%
\institute{University of Massachusetts Lowell,Lowell MA, USA \and
University of Massachusetts Chan Medical School, Worcester, MA, USA
\and
National Institute on Aging, National Institute of Health, Bethesda, MD, USA\\
\email{
\{Sharmin\_Sultana,Yidong\_Zhu\}@student.uml.edu,
\{ShaoHsien.Liu,MohammadArifUl.Alam1\}@umassmed.edu}}
\maketitle              
\begin{abstract}
The integration of diverse health data, such as IoT (Internet of Things), EHR (Electronic Health Record), and clinical surveys, with scalable AI(Artificial Intelligence) has enabled the identification of physical, behavioral, and psycho-social indicators of pain. However, the adoption of AI in clinical pain evaluation is hindered by challenges like personalization and fairness. Many AI models, including machine and deep learning, exhibit biases, discriminating against specific groups based on gender or ethnicity, causing skepticism among medical professionals about their reliability. This paper proposes a Multi-attribute Fairness Loss (MAFL) based Convolutional Neural Network (CNN) model designed to account for protected attributes in data, ensuring fair pain status predictions while minimizing disparities between privileged and unprivileged groups. We evaluate whether a balance between accuracy and fairness is achievable by comparing the proposed model with existing mitigation methods. Our findings indicate that the model performs favorably against state-of-the-art techniques. Using the NIH All-Of-US dataset, comprising data from 868 individuals over 1500 days, we demonstrate our model's effectiveness, achieving accuracy rates between 75\% and 85\%.

\end{abstract}
\section{Introduction}

The Intelligent Internet of Things (IIoT) is helping to improve healthcare by allowing for continuous monitoring of health vitals and behaviors, which can lead to just-in-time health interventions to the needful communities \cite{9,10,11}. Given the hype and potential to revolutionize health assessment through IIoT coupling with hospital records, heterogeneous data availability has emerged as the largest dimension than ever and revolutionizes better health care by improving all aspects of patient care, including safety, effectiveness, patient-centeredness, education, and efficiency\cite{6,175,176}. For all its potential, the use of Artificial Intelligence (AI) in healthcare also brings major risks and potential unintended harm by introducing biases in AI decisions. Such biased or unfair AI may discriminate its decisions based on persons' sensitive attributes \textcolor{red}{also known as protected attributes} (such as race, gender, ethnicity, and disabilities) which result in mistrust among clinicians and prevent AI systems to adopt in clinical settings \cite{12,13}. Despite efforts to close these gaps, racial/ethnic minorities continue to have poorer healthcare and outcomes \cite{intro2}. In AI, such discrimination is referred to as bias \cite{inprocessing}.

Bias indicates the inclination of judgment towards a certain group of people. Bias may arise from diverse origins, including unintentional errors by individuals, historical prejudices, and algorithmic parameters \cite{reason_bias}. Measuring bias is challenging because it may occur explicitly or implicitly\textcolor{red}{\cite{intro3}}. Explicit bias involves conscious attitudes that can be measured by self-report while implicit biases occur outside of conscious awareness and can result in a negative evaluation of a person based on \textcolor{red}{protected} attributes \cite{intro1}. \textcolor{red}{Protected} attribute refers to the features that partition the entire population into groups and these groups influence the parity of the outcome. {\it Privileged Group} refers to the group of people in the protected attributes who are at systematic advantage level, \textcolor{red}{generally having opportunities, benefits, or advantages in society that are not available to all groups. These privileges are often unearned and may be invisible to those who have them}. {\it Unprivileged Group} \textcolor{red}{is on the other side}, refers to the group of people in the protected attributes who are at a systematic disadvantage level.

In a clinical context, the most effective approaches to evaluating an individual's health status are those that utilize straightforward and comprehensible questions and procedures. Self-reported questionnaires such as the Iowa Pain Thermometer (IPT) and Numeric Rating Scale (NRS) excel in assessing pain severity. However, the task of accurately diagnosing and managing pain remains challenging for healthcare providers due to its widespread occurrence and negative implications \cite{20,21,22,23,24}. 

Additionally, the presence of racial bias in healthcare further complicates pain assessment and treatment. Medical professionals who endorse themselves false beliefs about biological differences between the two groups (Black or White) rate a Black patient's pain lower and provide less accurate treatment recommendations \cite{race_bias}. They also highlight existing research revealing disparities in pain treatment between Black and white patients, where Black patients are less likely to receive pain medications and are often given lower quantities. This concerning trend persists even in cases involving pediatric patients, revealing both overprescription for White patients and underprescription for Black patients. Furthermore, gender bias also plays a significant role in pain evaluation. A study reveals that female patients' pain tends to be underestimated due to ingrained gender stereotypes and biases \cite{gender_bias}. Hence it's critical to understand the limitations of specific assessment techniques as well as the significant influence that  biases have on the diagnosis and management of pain.

This paper introduces a research contribution centered on predicting patients' pain status using heart rate and step-count data obtained from wearable devices (Fitbit) and Electronic Health Records (EHR). The proposed approach utilizes a novel CNN method with Multi-attribute Fairness Loss (MAFL), incorporating all \textcolor{red}{protected} attributes (age, gender, race, ethnicity, cognitive ability) from the data and aiming to address disparities among \textcolor{red}{protected} attributes.  The effectiveness of the fair pain assessment model is evaluated using NIH All-Of-US data \cite{27}, analyzing a cohort of 868 individuals' wearable and EHR data collected over 1500 days. The proposed model aims to promote fairness in classification by reducing differences between privileged and disadvantaged groups based on these attributes.

\section{Literature Review}
A lot of extraordinary work has been reported in the literature on pain assessment for wearable devices. However, to the best of our knowledge, there has been very limited work reported that uses bias detection and mitigation techniques to predict pain status. Below are some notable works which we studied in detail to understand the challenges in this research domain.
\subsection{Clinical Pain Assessment Data Biases}
Research has shown that racial, age-related, and socioeconomic biases exist in pain treatment, with nonwhite and older patients and those of lower socioeconomic status being less likely to receive pain medication \cite{bias3,bias4}. Among others, age-related bias \cite{bias7}, (i.e., older adults being less likely to receive pain medications for pain compared to younger adults), socioeconomic bias \cite{bias8} (i.e., lower socioeconomic status influencing clinician decision-making in pain treatment), have been observed significantly in clinical settings. Beside that, data may get biased to the light-skin-toned people over dark skin people in the publicly available datasets \cite{subgroup}  whereas  performance ML (Machine Learning) models also affect because of gender, age, and racial attributes \cite{drliu}. Such biases skewed the decision of the models to the advantageous groups. 

\subsection{Wearable-based Pain Assessment Automation}
Wearable devices offer a promising opportunity to enhance pain assessment outcomes in clinical and research settings. The underlying concept is that pain triggers sympathetic physiological responses, reflected in increased heart and respiratory rates, blood pressure, and other measures' \cite{wearable4,wearable5}. While wearable-based pain assessment research is limited, a study showed that biweekly coaching with wearables improved physical activity in knee osteoarthritis patients, potentially alleviating pain \cite{wearable6}. Correlation studies have explored pain biomarkers through wearables; for instance, Hallman et al. linked reduced neck/shoulder pain in males with less work-related sitting time \cite{wearable7}, Jacobson and O'Cleirigh predicted pain levels (0-10) with 74.63\% accuracy using minute-level activities for HIV patients \cite{wearable8}, and Johnson et al. achieved 0.729 error rate in pain level prediction using Microsoft Band 2's accelerometer and heart rate data \cite{wearable9}.

\subsection{Bias in Wearables-based Health Vitals Monitoring System}
While earlier discussions hinted at wearables' potential for unbiased pain assessment, recent studies reveal that wearable sensors aren't immune to biases. These wearables offer a wealth of health-related data, encompassing heart rate anomalies, personal electrocardiogram (ECG) monitors, sleep trackers, and pulse pressure devices for a healthier lifestyle \cite{wearable10}. However, findings show that photoplethysmographic (PPG) green light signals lead to heart rate miscalculation in individuals with lighter skin tones, and galvanic skin response devices are less accurate for those with darker skin tones \cite{wearable11}. Despite limited coverage, the impact of dark skin tones on wearable accuracy remains a largely unaddressed issue necessitating industrial and scientific attention. Similarly, wearable-based health monitoring might engender disparities in older adults, and individuals with skin conditions, and possibly persist in wearable-based machine learning algorithms \cite{arif_fairness}.

In this paper, we focus on pain progression alignment with the help of wearable (Fitbit) and EHR data. To predict pain status, we introduce MAFL, a novel loss function-based CNN model that incorporates all \textcolor{red}{protected} attributes present in the dataset. The aim of our implemented model is to reduce disparities among privileged and unprivileged groups.  

\section{Proposed Methodology}
\begin{figure*}[h!]
\captionsetup{justification=centering}
\begin{center}
\includegraphics[width=\linewidth,scale=1.5]{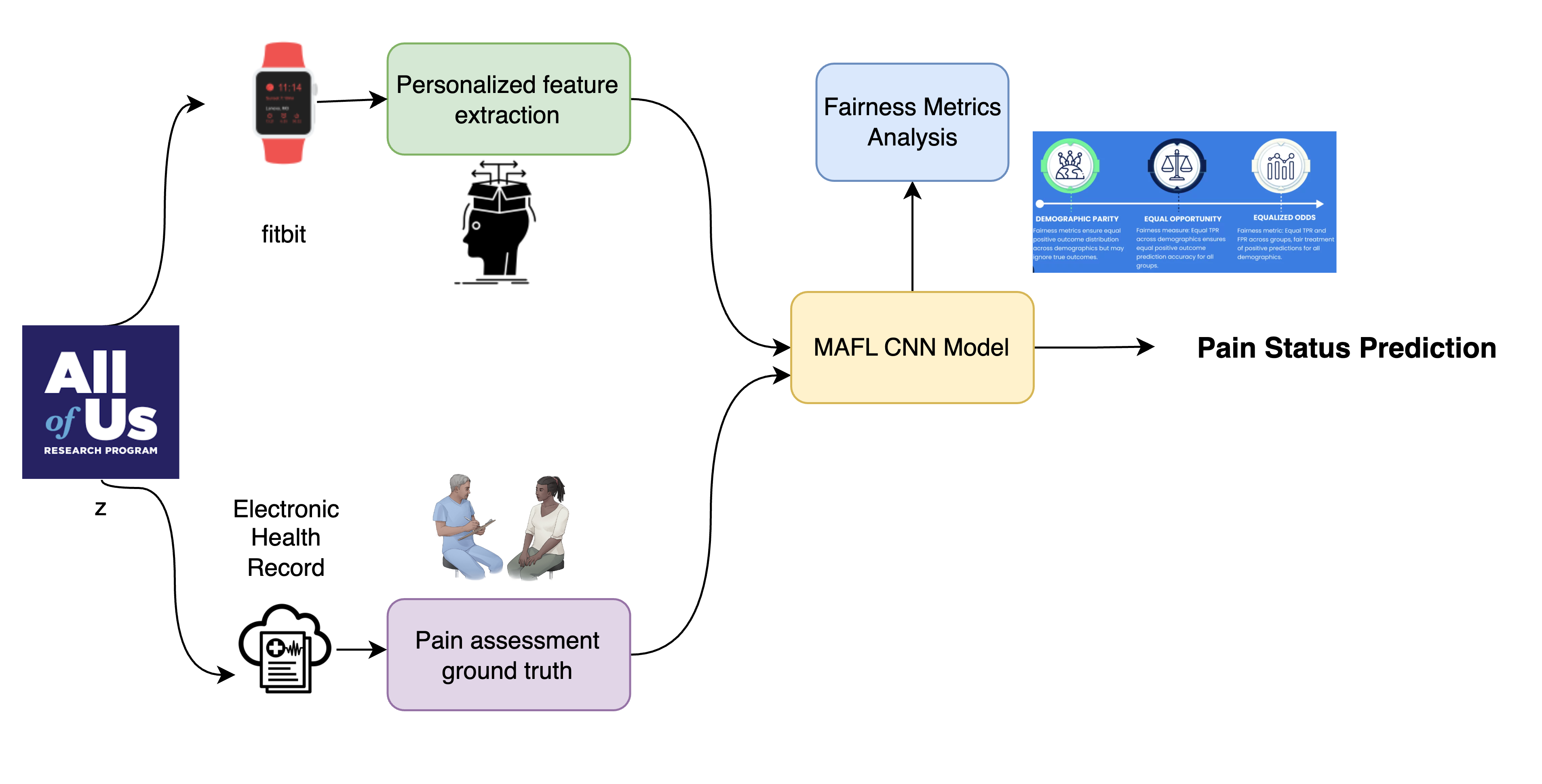}
\caption{Proposed Wearables and Electronic Health Record (EHR) based Fair Pain Assessment Automation Pipeline}
\label{fig:overview}
\end{center}
\end{figure*}

Our proposed framework consists of the following core modules, as depicted in Figure.\ref{fig:overview}:
\begin{enumerate}
    \item \textbf{Data Collection:}In this module, we utilize the expansive NIH All-Of-Us dataset to query and select wearable sensor data, specifically data from Fitbit devices, in conjunction with corresponding EHR data. This step forms the foundational data source for our pain assessment framework.
    \item \textbf{Time Series Feature Extraction:} This module focuses on the extraction of features from time-series data. 
    \item \textbf{Pain Assessment Ground Truth Extraction:} In this module, we perform ICD-10 code matching and interpretation to extract pain assessment ground truths from the All-Of-Us EHR database. These ground truths serve as the basis for training and validating our pain assessment model.
    \item \textbf{Bias Detection} Five fairness metrics were used to estimate the intensity of bias. Identifying bias within pain assessment is crucial to ensure fairness and equity in our framework.
     \item \textbf{Bias Mitigation:} To address identified biases, we introduce a Multi-attribute Fairness Loss (MAFL) based CNN model. This model is designed to detect pain levels using personalized features and pain assessment ground truth data. Its primary goal is to reduce disparities in pain assessment among privileged and underprivileged demographic groups.
\end{enumerate}

\subsection{NIH All-of-Us Data Collection}
To access NIH All of Us data, we registered and signed a Data Use and Registration Agreement (DURA). Each member underwent the Collaborative Institutional Training Initiative (CITI Program) training and applied for Controlled Tier access. To date, the All Of Us Research Program collected diverse data from 327,000+ participants without missing values with the help of 100+ funded partner organizations from 460+ sites where 50\% of them are racial and ethnic minorities. Among the entire data, 11,700 participants have wearable (Fitbit watch), EHR, and all forms of EMA (the basic, overall, health, lifestyle, pain assessment, personal medical history, etc.) surveys. The inclusion criteria of this paper are, that participants must have been evaluated with pain in the hospital visit at least twice in the same year and each of the wearable sensor values (Fitbit minute level step count and minute level average heart rate) should not have any missing values more than 10\%. Among these 11,700 population of the whole dataset, only 636 (5\%) match the inclusion criteria. We used the concept ID `3036453' from the EHR that provides a Visual Analog Score (VAS) of the pain assessment tool ranging from 0-10 where 10 is estimated as the highest level of pain and 0 is estimated as no pain.

\subsection{Time Series Feature Extraction}
Time series feature extraction serves as the initial stage in traditional machine learning methods. We analyze minute-level total step count and average heart rate data from the Fitbit wearable sensor, extracting more than 60 diverse features across temporal, statistical, and spectral domains for each individual data point. Table \ref{tab:extracted_features} shows the details of our extracted features in the above-considered domain.
\begin{table}[!htbp]
 \begin{center}
 \caption{Statistical, temporal, and spectral domain feature details}
 \begin{tabular}{|p{3.2in}|}
 \hline
{\bf Statistical domain features}\\
 \hline
   Empirical Cumulative Distribution Function (ECDF), ECDF Percentile, ECDF Percentile Count, Histogram, Interquartile range, Kurtosis, Max, Mean, Mean absolute deviation, Median, Median absolute deviation, Min, Root mean square, Skewness, Standard deviation, Variance
\\ \hline
\ {\bf Temporal domain features}\\
 \hline
 
Absolute energy, Area under the curve, Autocorrelation, Centroid, Entropy, Mean absolute diff, Mean diff, Median absolute diff, Median diff, Negative turning points, Peak-to-peak distance, Positive turning points, Signal distance, Slope, Sum absolute diff, Total energy, Zero crossing rate, Neighbourhood peaks
\\ \hline
\ {\bf Spectral domain features}\\
 \hline
 
FFT mean coefficient, Fundamental frequency, Human range energy, linear prediction cepstral coefficients (LPCC), Mel-frequency cepstral coefficients (MFCC), Max power spectrum, Maximum frequency, Median frequency, Power bandwidth, Spectral centroid, Spectral decrease, Spectral distance, Spectral entropy, Spectral kurtosis, Spectral positive turning points, Spectral roll-off, Spectral roll-on, Spectral skewness, Spectral slope, Spectral spread, Spectral variation, Wavelet absolute mean, Wavelet energy, Wavelet standard deviation, Wavelet entropy, Wavelet variance
\\ \hline
 \end{tabular}
 \label{tab:extracted_features}
 \end{center}
 \vspace{-5mm}
\end{table}

In this paper, we did personalized feature extraction where only the total step count and average heart rate every minute are considered in our data set. We use the feature deviance method to extract personalized features for each individual. In this regard, at first, we extract statistical, temporal, and spectral domain feature $X^{fi}_d$ where $f$, $i$ and $d$ refer to the type of feature $f\in F = \{statistical, temporal, spectral\}$, index of feature $i \in N=\{1,2..N\}$ (we have total N=60 features), day of the collected wearable data. We consider 4 different deviant functions on these 60 extracted features described as follows:
\begin{enumerate}
    \item {\bf Mathematical deviance:} This function directly subtracts previous day's features $X^{fi}_{d-1}$ from current day's, $X^{fi}_d$.
    \begin{equation}
        X^{fi}_d(1)=X^{fi}_d-X^{fi}_{d-1}
        \label{eqn:deviance}
    \end{equation}
    \item {\bf Logarithmic deviance:} This function subtracts the previous day's logarithmic features from the current day's logarithmic features.
    \begin{equation}
        X^{fi}_d(2)=log(X^{fi}_d)-log(X^{fi}_{d-1})
    \end{equation}
    \item {\bf Cosine deviance:} This function multiplies the previous day's cosine features with the current day's cosine features.
    \begin{equation}
        X^{fi}_d(3)=cos(X^{fi}_d) * cos(X^{fi}_{d-1})
    \end{equation}
    \item {\bf LogCosh deviance:} This function can be defined as the following equation.
    \begin{equation}
        X^{fi}_d(4) = log(\frac{(e^{x} + e{-x})}{2})
    \end{equation}
    where x is the mathematical deviance $X^{fi}_d(1)$ from Equation \ref{eqn:deviance}.
\end{enumerate}
Finally, we normalize all of the features thus they can be fed into the model.
\subsection{Pain Assessment Ground Truth Extraction}
These days, clinicians are more concerned about the changes (recovering or worsening) of pain rather than pure pain assessment with a scale ranging from 0 to 10, \cite{bias1,bias2}, we consider pain assessment as a pain recovery tracking progression (recovering or worsening than last assessment) to pace with the modern clinical pain progress tracking practice. In this regard, at first, we consider only the days when we have a ground truth of pain assessment using clinically validated tools extracted from EHR. Then, consider our problem as a binary classification, where the pain recovery will be labeled as `True' if a patient's current state of pain is improved (pain level score gets reduced); and `False' otherwise. To accommodate such pain recovery tracking problem, we require at least one prior pain assessment conducted by clinically validated tools. Such clinician/nurse evaluated prior pain assessment comparisons to enhance the `human accountability' which is a core component of ML fairness \cite{zhou21}.

\subsection{Bias Detection}
In this paper, we used 5 fairness metrics to check bias in our data and mitigation algorithm. Here, we use 'Unprv' to refer to the unprivileged group and 'Prv' to refer to the privileged group in all equations.

{\bf Statistical Parity Difference} is the difference between the favorable outcome for unprivileged groups and privileged groups. Dwork et al.(2012) introduced the concept as "statistical parity" or "demographic parity" and defined it mathematically as below equation where A is the protected attribute \cite{DworkFairness2012}, in our case the 'Prv'. The ideal value is 0 and this metric is considered fair for values ranging from -0.1 to 0.1 . 
\begin{equation}
    P(Y=1|Unprv) - P(Y=1|Prv)
\end{equation}

{\bf Disparate Impact} refers to the ratio of favorable outcomes for the unprivileged group to the privileged group. The ideal value for disparate impact is 1 and it is considered fair if the value ranges between 0.8 to 1.25. If the disparate impact value is less than 1, it gives benefits to the privileged group otherwise unprivileged group gets the advantage. Feldman et al.(2015) introduced the 80 percent rule from US labor law into algorithmic fairness and formalized the DI \cite{FeldmanDisparate2015}, in our case which can be written as:

\begin{equation}
    \frac {P(Y=1|Unprv)}{P(Y=1|Prv)}
\end{equation}

{\bf Equal Opportunity Difference} computes the  true positive rate difference between unprivileged and privileged groups whereas the true positive rate is the ratio of true positives to the total number of actual positives for a given group. The ideal value is 0 and the fairness range for this metric is -0.1 to 0.1. If the value is greater than 0, unprivileged groups are considered to benefit and if it is less than 0, privileged groups are the beneficiary. Hardt et al.(2016) introduced equal opportunity as TPR parity between groups and defined it mathematically \cite{HardtEquality2016}, in our case can be presented as: 
\begin{equation}
    TPR = \frac{\# \text{True positive predictive values}}{\text{Total positive values}}
\end{equation}

\begin{equation}
        TPR(Unprv) - TPR(Prv)
\end{equation}

{\bf Theil Index} is a statistical measure of inequality that computes the distance away from the ideal value. The perfect value is 0, the lower value represents fairer metrics while the higher value indicates a problem. Speicher et al.(2018) adapted Theil index from economics to measure machine learning fairness and showed how it captures both individual and group unfairness \cite{SpeicherUnified2018}.

{\bf Average Odd Difference} is the average difference between the false positive rate and true positive rate between unprivileged and privileged groups. Hardt et al. (2016) introduced equalized odds, which average odds difference is based on and defined it using both FPR and TPR differences \cite{HardtEquality2016}, the formula in our case is 

\begin{equation}
      \frac{(FPR_{Unprv} - FPR_{Prv}) + (TPR_{Unprv} - TPR_{Prv})}{2}
\end{equation}
While the perfect value of this metric is 0, it is considered fair if the value ranges between -0.1 to 0.1. A value less than 0 implies benefit to the privileged group otherwise unprivileged group takes advantage.

We focused on Statistical Parity Difference (SPD) and Disparate Impact (DI) for initial bias detection due to their complementary nature \cite{MehrabiFairness2021}. SPD provides an absolute difference measure, while DI offers a ratio-based assessment, allowing us to capture both absolute and relative disparities in outcomes. For comprehensive evaluation, we employ all these five fairness metrics.

\subsection{Bias Mitigation}
Pain assessment is frequently influenced by a patient's demographic information, resulting in unequal care for certain groups. Hence, it's vital to establish a system free from bias to ensure equitable treatment for all segments of the population. In this paper, we designed a loss function MAFL that incorporates fairness constraints (demographic parity) alongside traditional binary cross-entropy loss. Then we train the CNN model with MAFL on our dataset. During training, the model will learn to minimize the disparities between privileged and unprivileged groups. 
\subsubsection{Design of FairLoss} \

\fbox{\begin{minipage}{0.95\linewidth}
    \footnotesize
    \begin{tabular}{l}
    
    \textsc{\textbf{Multi\_Attribute\_Fair\_Loss}}($y_{\text{true}}, y_{\text{pred}}$): \\
    \hspace{1em} \textit{unprivileged\_indices} $\gets$ indices where \textit{sensitive\_attributes} equals 0 \\
    \hspace{1em} \textit{privileged\_indices} $\gets$ indices where \textit{sensitive\_attributes} equals 1 \\

    \hspace{1em} \textit{loss\_unprivileged} $\gets$ mean of $y_{\text{pred}}$ values at \textit{unprivileged\_indices} \\
    \hspace{1em} \textit{loss\_privileged} $\gets$ mean of $y_{\text{pred}}$ values at \textit{privileged\_indices} \\
    \hspace{1em} \textit{disparity} $\gets$ \textit{loss\_privileged} - \textit{loss\_unprivileged} \\
    
    \hspace{1em} $\lambda \gets 1.0$ \\
    \hspace{1em} \textit{loss} $\gets$ Binary\_Cross\_Entropy($y_{\text{true}}$,$y_{\text{pred}}$) \\
    \hspace{1em} \textit{loss} += $\lambda \cdot \textit{disparity}^2$ \\
    \hspace{1em} \textit{regularization\_loss} $\gets$ 0.01 $\cdot$ $ \left \lvert \sum_{k=1}^{n} (y_{\text{pred}} - mean y_{\text{pred}} ) \right \rvert $ \\
    \hspace{1em} \textit{total\_loss} $\gets$ \textit{loss} + \textit{regularization\_loss} \\

    \hspace{1em} \textbf{return} \textit{total\_loss}
    \\
    \end{tabular}
\end{minipage}}

\vspace{0.2cm}

\begin{figure*}[!htb]
\begin{center}
 \includegraphics[width=\linewidth, height=5cm, scale=1.5]{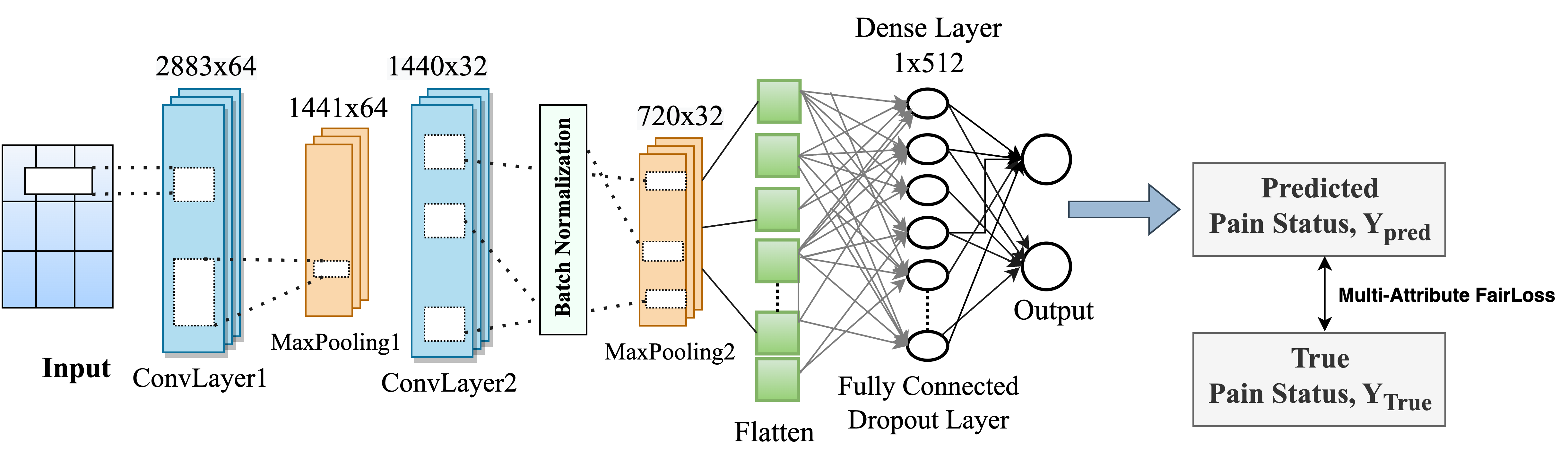}
 \caption{The architecture of 1D deep Convolutional Neural Network (CNN) for pain level assessment}
 \label{fig:cnn}
\end{center}
\end{figure*}

MAFL adds fairness to the regular loss function. Initially, it takes the indices of the unprivileged and privileged groups within the sensitive attributes list and then calculates the mean prediction values for the unprivileged and privileged groups separately. The difference represents the disparity between the groups in terms of the predictions made by the model. A regularization term is added to the loss function to further promote fairness. The term penalizes large differences between individual predictions and the mean prediction of all samples. Finally, the total loss is calculated by summing the regular loss function (in this case, it is Binary Cross-Entropy) and the regularization loss. Here, a fairness score is applied to input and target for sensitive attributes (k=1 to n). In our dataset, we have 5 protected attributes (gender, race, ethnicity, age, and cognitive ability (dementia)).  MAFL avoids using sensitive attributes instead, focuses on other relevant factors (average heart rate and step count) to make predictions.

\subsubsection{Integrated CNN Framework to Predict Pain Status} \

Figure \ref{fig:cnn} depicts the architecture of the deep convolution neural network for classification. It is made up of seven layers: an input layer, two 1D-convolutional layers, two max-pooling layers, one flattening layer, one fully connected hidden layer, and an output layer. To predict the pain level, the softmax function is applied to the nodes in the output layer. The input layer contains 2883 input numbers that represent demographic features as well as heart rate and step counts for a day. The hidden features generated by the first convolution layer's 64 filters with window size three are used as input for the second convolution layer, which transforms them in the same way via convolution, batch-normalization, and non-linear transformation of its inputs with the rectified-linear unit (ReLU) activation function. Following that, a 1D max-pooling layer is added to convert the variable length hidden features in the previous convolution layer to a fixed number of features. The extracted features are then fed into a 512-node fully-connected hidden layer. To predict the probability of pain level, these 512 nodes are fully connected to two nodes in the output layer. The softmax activation function is used by the node in the output layer. The dropout technique is used in the hidden layer to prevent over-fitting (i.e., the 6th layer in Fig. \ref{fig:cnn}).

\section{Experiment Evaluation}

\subsection{Pre-process the Dataset}
For any classification model to work, raw data must be converted into a clean data set, which means the data set must be converted to numeric data. In the beginning, we converted the string type date attribute to python supported DateTime module and then compared the previous date's pain with the next date's pain. If pain improves then we assign the pain level as 1 otherwise 0. After that, we included cognitive ability (dementia) column for each person based on the list of patients who have dementia. Numeric data conversion was accomplished by encoding all categorical features as binary column vectors. Since data on a person's heart rate and step counts for a day were in list format, we split the data into multiple columns (1440 columns for heart rates and 1440 columns for step counts for each day). If any rows are missing, we fill them with mean or interpolated values. Finally, we dropped columns that are not necessary to carry out our work after re-scaling them using the Min-Max normalization technique.

\subsection{Protected attribute selection}

The clean dataset we used, has 868 records. In our dataset, we have 5 protected attributes such as gender, race, ethnicity, age, and cognitive ability (dementia). \textcolor{red}{The term "race" and "ethinicity" are often used interchangeably in everyday language, but they refer to different concepts in social science and identity studies. Banton\cite{Banton2015} defines race as a category based on physical characteristics that are perceived as significant by society. The concept of race is seen as a way to categorize and often hierarchize individuals and groups based on superficial traits. While, ethnicity refers to the cultural factors that distinguish groups of people from one another. Ethnicity is viewed as a source of identity and belonging, often linked to cultural and social practices rather than physical characteristics}.

\begin{figure}[!htb]
\begin{minipage}[b]{0.45\linewidth}
\begin{center}
   \includegraphics[width=\textwidth]{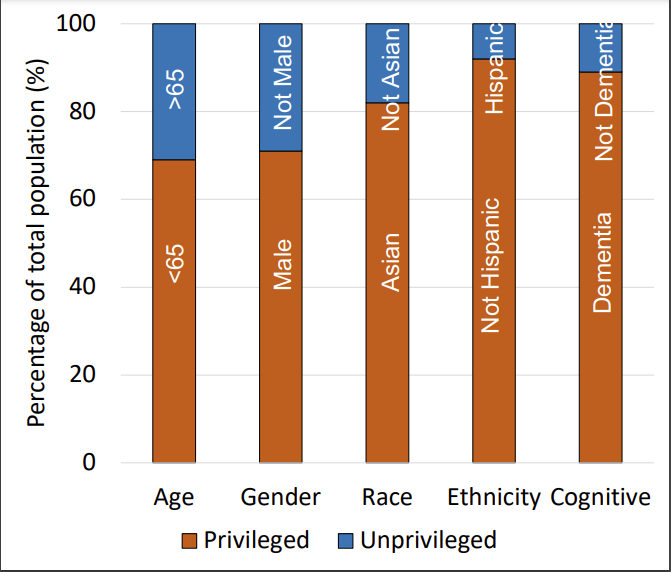}
    \caption{Rich Subgroups for Each Protected Attribute}
    \label{fig:protected_attribute} 
\end{center}
\end{minipage}
\hspace{0.2cm}
\begin{minipage}[b]{0.45\linewidth}
\begin{center}
   \includegraphics[width=\textwidth]{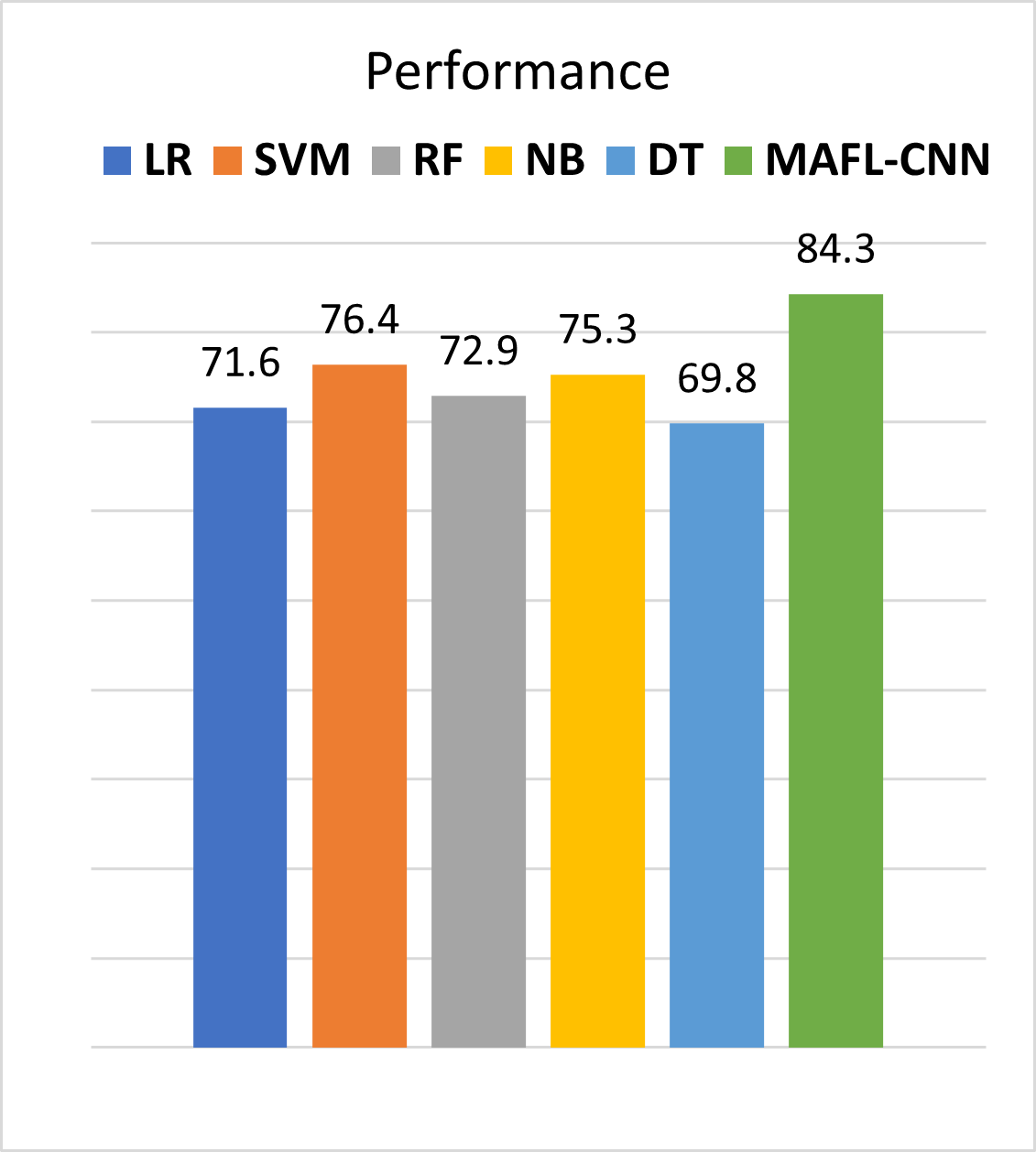}
    \caption{Pain Assessment Performance for Classification Models}
    \label{fig:accuracy_ml} 
\end{center}
\end{minipage}
\end{figure}

To select privileged groups for each sensitive attribute, we determined which groups are rich subgroups i.e. who are most advantageous than others \cite{subgroup}. \textcolor{red}{According to National Institutes of Health(NIH)\cite{NCBIAgeGroups1998,NIHAgeGuidelines2024} adults 65 or older are generally described as "elderly"}.
For our dataset, we determined that males, Asians, not Hispanic or Latin, age less than 65, and people with no dementia are the rich subgroups
for each protected attribute respectively which are called gender, race, ethnicity, aging, and cognitive disability biases respectively (Fig. \ref{fig:protected_attribute}).

\subsection{Implementation Details}
To evaluate our experiment, we conduct tests on our dataset using benchmark pre-processing, in-processing, and post-processing bias mitigation techniques suggested by AIF360\cite{aif360} to show that MAFL-CNN satisfies the tradeoff between fairness and accuracy. To begin, we transformed our dataset into Standard Dataset format and then partitioned it randomly by 80-20\%; 80\% training data, and 20\% test data for each protected attribute. We calculated the dataset metrics i.e. statistical parity difference and disparity impact ratio for both training and test datasets. Then we assess the effectiveness of our MAFL-CNN model in terms of fair accuracy by comparing it to cutting-edge fairness-aware AIF360 bias mitigation techniques on Logistic Regression (LR), Random Forest (RF), Decision Tree (DT), Support Vector Machine (SVM), Naive Bayes (NB), and CNN models. For the \textit{logistic regression}, we use liblinear solver and set the maximum iteration at 50,000. In \textit{random forest}, we set maximum depth value 2. For the \textit{decision tree} classifier, the entropy criterion is used to find the optimal split where entropy represents the disorder of features with respect to the target class. We used balanced class weight for \textit{support vector machine} classifier that helps to reduce execution time. In \textit{Naive Byes (NB)}, we use the Bernoulli NB model while binary cross entropy (BCE) was used as a loss function in the CNN model. For our \textit{\textbf{MAFL CNN}} model, we set the two filter sizes (i.e. 64 and 32) for the convolution layers and used 512 fully connected dense layers. We run our model for 20 epochs. We selected the best model based on the highest number of fair classification matrices (Statistical Parity Difference, Disparate Impact, Equal Opportunity Difference, Theil Index, and Average Odd Difference) to satisfy accuracy and fairness trade-off.

To carry on the whole experiment, a virtual environment has been created using python3.11. Our CPU configuration was Intel Xeon CPU (2.00GHz) processor with 12 GB of RAM and the GPU was Tesla K40c with 12 GB of RAM. It takes approximately 170 minutes to complete the task.

\section{Results}
\subsection{Dataset Bias Detection and Mitigation}

\begin{figure}[!htb]
\begin{center}
 \includegraphics[width=0.9\linewidth]{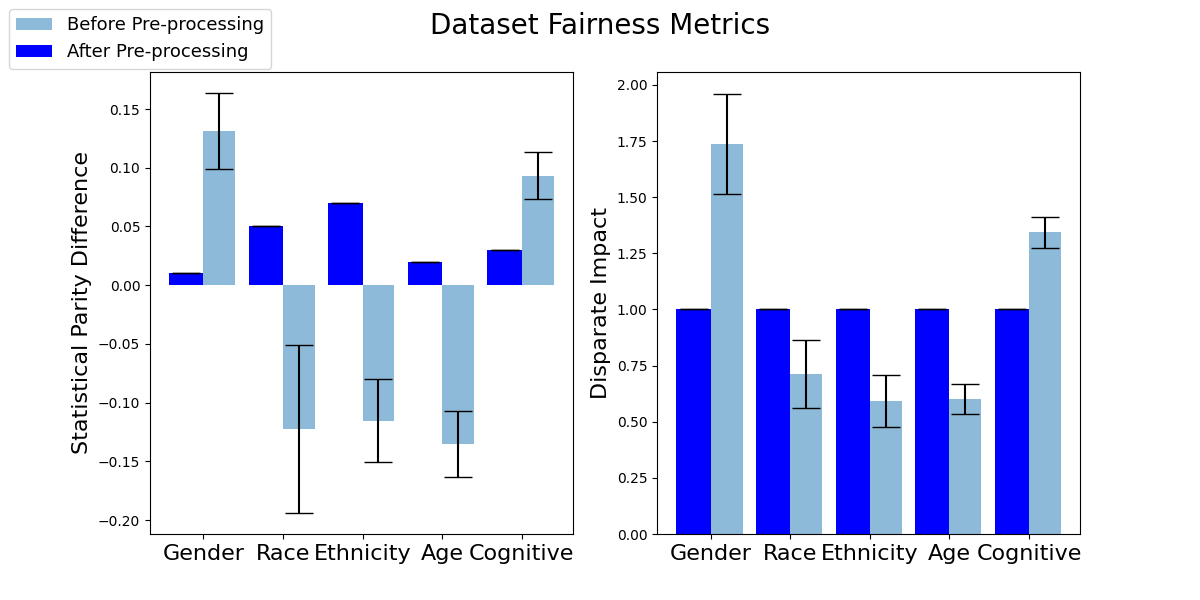}
 \caption{Statistical Parity Difference (SPD) and Disparate Impact (DI) before and after applying pre-processing algorithms for different protected attributes. The black bars indicate the extent of $\pm 1$ standard deviation. The ideal fair value of SPD is 0 and DI is 1.
}
 \label{fig:disparity}
\end{center}
\end{figure}

To assess the presence of bias in our dataset, we employed two fairness metrics, namely statistical parity difference and disparate impact, as described in the work by AIF360 \cite{aif360}.  If the unprivileged group receives a positive outcome of less than $80\%$ of their proportion relative to the privileged group, it suggests a potential bias or disparate impact. In our original dataset, prior to any mitigation efforts, the disparate impact value was $0.72$, and the statistical parity difference was $1.5$, which ideally should range between $-1$ and $1$.  So clearly it indicates that our dataset is compromised in decision making. To address these biases, we made modifications to the original dataset using fair pre-processing procedures\cite{aif360}:\\
Reweighing:
We first assign weights to balance representation across(group, label) combinations, then maintain features while adjusting instance weights to ensure fairness \cite{aif360}.
Disparate Impact Remover:
We use a repair level parameter(set to 1.0) to control the strength of the transformation and transform features to remove discriminatory information while preserving rank-ordering within groups, only modifies the features, labels or protected attributes are not affected \cite{aif360}.\\
To evaluate the impact of these pre-processing bias mitigation techniques on bias detection metrics, we computed fairness metrics both before and after the transformation. The results, shown in Figure \ref{fig:disparity}, indicate the extent of $\pm 1$ standard deviation using the bars. It is evident that the Reweighing and Disparate Impact Remover pre-processing techniques enhance fairness for all the datasets in terms of both provided measures.

\subsection{Performance of Mitigation Algorithm}
To evaluate the efficacy of our mitigation algorithm, MAFL-CNN, we perform a comparative assessment of its classification accuracy against a selection of advanced mitigation algorithms provided by IBM, as detailed in \cite{aif360}. As part of AIF360 mitigation techniques, we employed six state-of-the-art classification models, namely, Support Vector Machine (SVM), Random Forest (RF), Logistic Regression (LR), Decision Tree (DT), Naive Bayes (NB), and Convolutional Neural Network (CNN). Figure \ref{fig:accuracy_ml} illustrates the collective performance of the classification models. It is evident from the chart that MAFL-CNN consistently outperforms all other models, showcasing its superiority as a fairer classification technique. 
\begin{table}[!htbp]
\begin{scriptsize}
\vspace{-.2in}
\caption{ \scriptsize Bias detection and mitigation results for sensitive attributes: Age, Gender, Race, Cognitive Ability (Dementia), and Ethnicity. We compare MAFL-CNN with 5 state-of-art ML models and 1 NN model: SVM, RF, LR, DT, NB, and CNN referring to Support Vector Machine, Random Forest, Logistic Regression, Decision Tree, Naive Bayes, and Convolution Neural Network. We also evaluate MAFL-CNN with respect to AIF360 mitigation techniques (RW, DIR, EGR, AD, PR, and ROC referring to Reweighing, Disparate Impact Remover, Exponentiated Gradient Reduction, Adversarial debiasing, Prejudice Remover techniques, and Reject Option Classification).}

\begin{center}
\begin{tabular}{|p{1.8cm}|p{.75cm}|p{.75cm}|p{.75cm}|p{.75cm}|p{.75cm}|p{.75cm}|p{.75cm}|p{.75cm}|p{.75cm}|p{.75cm}|p{.75cm}|}
    \hline
    \multirow{2}{*}{Mitigation} & \multicolumn{2}{c|}{Age} & \multicolumn{2}{c|}{Gender}  & \multicolumn{2}{c|}{Race} & \multicolumn{2}{c|}{Cognitive} & \multicolumn{2}{c|}{Ethnicity} & \multicolumn{1}{c|}{Altogether} \\ 
    \cline{2-12}
    & Model & Acc \% & Model & Acc \% & Model & Acc \% & Model & Acc \% & Model & Acc \% & Acc \% \\ 
    \hline \hline

    \multicolumn{12}{|c|}{Before Bias Mitigation} \\
    \hline
    Before Mitigation 
    & LR & 82.0
    & CNN & 78.74
    & LR & 74.1
    & RF & 70.4 
    & LR & 78.3
    & -- \\
    \hline \hline

    \multicolumn{12}{|c|}{MAFL-CNN Mitigation} \\
    \hline
    \textbf{MAFL-CNN}
    & - & 77.3
    & - & \colorbox{yellow}{\textbf{75.2}}
    & - & 75.6
    & - & \colorbox{yellow}{\textbf{79.5}}
    & - & 78.7
    & \colorbox{yellow}{\textbf{84.3}} \\
    \hline \hline

    \multicolumn{12}{|c|}{AIF360 Mitigation} \\
    \hline
    RW - Preprocessing 
    & SVM & 76.2
    & \colorbox{yellow}{\textbf{\color{red} SVM}} & \colorbox{yellow}{\textbf{\color{red} 76.4}}
    & LR & 69.8
    & LR & 70.4 
    & CNN & 79.8
     & -- \\
    \hline

    DIR-Preprocessing
    & RF & 71.6 
    & LR & 67.6
    & DT & 69.8
    & RF & 70.6
    & \colorbox{yellow}{\textbf{\color{red} CNN}} & \colorbox{yellow}{\textbf{\color{red} 72.9}}
     & -- \\
    \hline

    EGR-Inprocessing
    & \colorbox{yellow}{\textbf{\color{red} LR}} & \colorbox{yellow}{\textbf{\color{red} 71.6}}
    & SVM & 65.7
    & SVM & 72.7 
    & SVM & 62.2 
    & \colorbox{yellow}{\textbf{\color{red} RF}} & \colorbox{yellow}{\textbf{\color{red} 73.4}}
    & -- \\
    \hline

    AD-Inprocessing
    & -- & 75.8 
    & -- & \colorbox{yellow}{\textbf{74.7}}
    & -- & 76.9
    & -- & 74.7 
    & -- & 71.3
    & -- \\
    \hline

    PR-Inprocessing
    & - & \colorbox{yellow}{\textbf{76.4}}
    & - & 75.7
    & - & 82.6
    & - & 77.1 
    & - & 72.5
    & --\\
    \hline

    ROC-Postprocessing 
    & - & 73.3
    & - & 75.2
    & - & \colorbox{yellow}{\textbf{80.6}}
    & - & 77.5 
    & - & 72.3
    & --\\
    \hline
\end{tabular}
\end{center}

\label{tab:individual_mitigation_result}
\end{scriptsize}
\end{table}

We present a comprehensive summary of the classification model results in Table \ref{tab:individual_mitigation_result}. In this table, we exclusively highlight the optimal model, characterized by its superior accuracy and reduced bias, for each mitigation technique. Our analysis focuses on sensitive attributes such as gender, age, race, ethnicity, and cognitive ability, and their impact on classification fairness. It is evident from the table that, across these attributes, all state-of-the-art machine learning models exhibit unfairness in terms of bias detection metrics. However, MAFL-CNN stands out as a promising performer, displaying consistently strong performance across all individual or combined sensitive attributes.
In the case of gender, cognitive ability, and $n$ sensitive attributes, MAFL-CNN surpasses other techniques by exhibiting superior performance in more than three fairness metrics. For age, ethnicity, and race, while the number of fairness metrics with superior performance is three or less, it remains noteworthy that MAFL-CNN consistently achieves higher accuracy compared to alternative mitigation methods.\\
Among the pre-processing techniques, SVM demonstrates the highest balance between accuracy and fairness metrics, achieving an accuracy of $76.4\%$. In the case of in-processing techniques, the Prejudice Remover (PR) technique performs well in terms of accuracy for the sensitive attribute 'Age', while the RF model shows fairness as a predictor for the Exponentiated Gradient Reduction (EGR) technique. The Reject Option Classification (ROC) technique exhibits an average accuracy of around $75\%$ for each sensitive attribute. However, our MAFL-CNN model suppresses the performance of all the models, resulting in accuracy ranging from $ 75\%$ to $85\%$.  Furthermore, unlike the AIF360 mitigation techniques, our MAFL-CNN model can consider multiple sensitive attributes. Additionally, it is worth noting that AIF360 only includes one neural network (NN) model, which is the adversarial debiasing algorithm, and its performance is noticeably lower than our MAFL-CNN model. So we can say that MAFL-CNN is successful in balancing classification accuracy and fairness.

\begin{sidewaysfigure}[!h]
\begin{center}
 \includegraphics[width=\linewidth]{Classification_Metrics_gender.png}
 \caption{Fairness vs. Accuracy before (top panel) and after (bottom panel) applying various bias mitigation algorithms. Five different fairness metrics are shown. The ideal fair value of disparate impact is 1, whereas for all other metrics it is 0. The circles indicate the mean value and the bars indicate the extent of $\pm 1$ standard deviation. \textcolor{red}{Protected} attributes: gender, race, age, ethnicity, dementia.}
 \label{fig:classification_metrics_all}
\end{center}
\end{sidewaysfigure}

We report the bias detection and mitigation results more elaborately in Fig \ref{fig:classification_metrics_all} with details of bias metrics and accuracy changes for all algorithm-bias mitigation pairs. Here, we visualize a sample outcome of the bias mitigation process (before and after mitigation) to understand the impact of the bias mitigation technique on accuracy and bias detection metrics. The top panel of this fig depicts the classification metrics before bias mitigation for all classifiers and the bottom panel shows the metrics after using mitigation techniques. We can clearly see that bias mitigation techniques significantly change the bias metrics estimations as well as the accuracy.

\section{Conclusion}
This paper presents a cutting-edge model that not only provides accurate predictions of patients' pain status but also contributes significantly to the advancement of fairness and equity in decision-making systems. Our experimental results demonstrate the efficacy of our proposed approach. We observe significant reductions in disparity between privileged and unprivileged groups while maintaining competitive classification accuracy [Figure \ref{fig:disparity}]. Furthermore, we analyze the impact of different \textcolor{red}{protected} attributes on the fairness metrics to gain insights into their influence on the classification process. By mitigating disparity in classification outcomes, our approach contributes towards promoting fairness and reducing bias in decision-making systems [Figure\ref{fig:accuracy_ml}] with accuracy ranging from $75\%$ to $85\%$ for existing \textcolor{red}{protected} attributes. This substantial accuracy range underscores the robustness and reliability of our proposed model. In summary, we have highlighted the current challenges and possible solutions to assess the pain automatically and ensure fairness. This work opens paths for future studies emphasizing exploring the generalizability of our method across different datasets and extending the approach to other deep learning architectures.

\end{document}